\documentclass[amssymb]{revtex4}
\usepackage{amsmath,graphicx}
\usepackage{epsfig}

\begin{document}

\title{Polymer desorption under pulling -\\ a $1^{st}-$order phase
transition without phase coexistence}

\author{\underline {A. Milchev}$^{1,2}$, V. G. Rostiashvili$^1$, S.
Bhattacharya$^1$, and
T.A. Vilgis$^1$}
\affiliation{$^1$ Max Planck Institute for Polymer Research, 10 Ackermannweg,
55128 Mainz, Germany\\
$^2$ Institute for Physical Chemistry, Bulgarian Academy of Sciences, 1113
Sofia, Bulgaria
}

\begin{abstract}
We show that when a self-avoiding polymer chain is pulled off a sticky surface
by force applied to the end segment, it undergoes a first-order thermodynamic
phase transition albeit {\em without} phase coexistence. This unusual feature is
demonstrated analytically by means of a Grand Canonical Ensemble (GCE)
description of adsorbed macromolecules as well as by Monte Carlo simulations of
an off-lattice bead-spring model of a polymer chain.

Theoretical treatment and computer experiment can be carried out both in the
{\em constant-force} statistical ensembl whereby at fixed pulling force $f$ one
measures the mean height $\langle h\rangle$ of the chain end above the adsorbing
plane, and in the {\em constant-height} ensemble where for a given height $h$
one monitors the resulting force $\langle f \rangle$ applied at the last
segment. We find that the force-assisted desorption undergoes a first-order
dichotomic phase transition whereby phase coexistence between adsorbed and
desorbed states does not exist. In the $f$-ensemble the order parameter (the
fraction of chain contacts with the surface) is characterized by huge
fluctuations when the pulling force attains a critical value $f_D$. In the
$h$-ensemble, in contrast, fluctuations are always finite at the critical
height $h_D$.

The derived analytical expressions for the probability distributions of the
basic structural units of an adsorbed polymer, such as loops, trains and tails,
in terms of the adhesive potential $\epsilon$ and $f$, or $h$, provide a full
description of the polymer structure and behavior upon force-assisted
detachment. In addition, one finds that the hitherto controversial value of the
universal critical adsorption exponent $\phi$ depends essentially on the extent
of interaction between the loops adsorbed chain so that $\phi$ may vary within
the limits $0.39 \le \phi \le 0.59$.

\end{abstract}
\maketitle

\section{Introduction}\label{intro}
Over the past decade, experimental force spectroscopy techniques such as Atomic
Force Microscopy (AFM) and optical or magnetic tweezers emerged as novel
methods which allow the manipulation of individual polymers with spatial
resolution in the {\em nm} range and force resolution in the {\em pN}
range\cite{Rief,Bustamante}. One can thus study the mechanical
properties and characterize the intermolecular interactions of a single
macromolecule which leads to better understanding of
the material elasticity on a molecular level\cite{Strick,Celestini}, enables
measuring the receptor - ligand binding strength\cite{Florin}, or the
determination of friction-induced energy dissipation during the movement of a
macromolecule on a solid surface\cite{Serr}.

The rapid development of experimental techniques has been followed by
theoretical considerations, based on the mean - field approximation
\cite{Sevick}, which provide important insight into the mechanism of polymer
detachment from adhesive surfaces under external pulling force. A comprehensive
study by Skvortsov {\it et al.} \cite{SKB} examines the case of a Gaussian
polymer chain. One should also note the close analogy between the forced
detachment by pulling and the unzipping of a double - stranded DNA.
Recently, DNA denaturation and unzipping have been treated by Kafri {\it et al.}
\cite{Kafri} using the Grand Canonical Ensemble (GCE) approach \cite{Poland,Bir}
as well as Duplantier's analysis of polymer networks of arbitrary topology
\cite{Duplantier}. An important result concerning the properties of adsorbed
macromolecule under pulling turns to be the observation \cite{Kafri} that the
universal exponents (which govern polymer loops statistics) undergo
renormalization when excluded volume effects between chain segments are taken
into account. In this work we use similar methods to describe the structure and
detachment of a polymer chain from a sticky substrate under pulling and
demonstrate the unusual properties of this phase transformation in two
conjugated statistical ensembles.

\section{Theory of chain desorption}\label{theory}

\subsection{A simplified case of detachment}

\begin{figure}[htb]
\hspace{-0.50cm}
\includegraphics[scale=0.25]{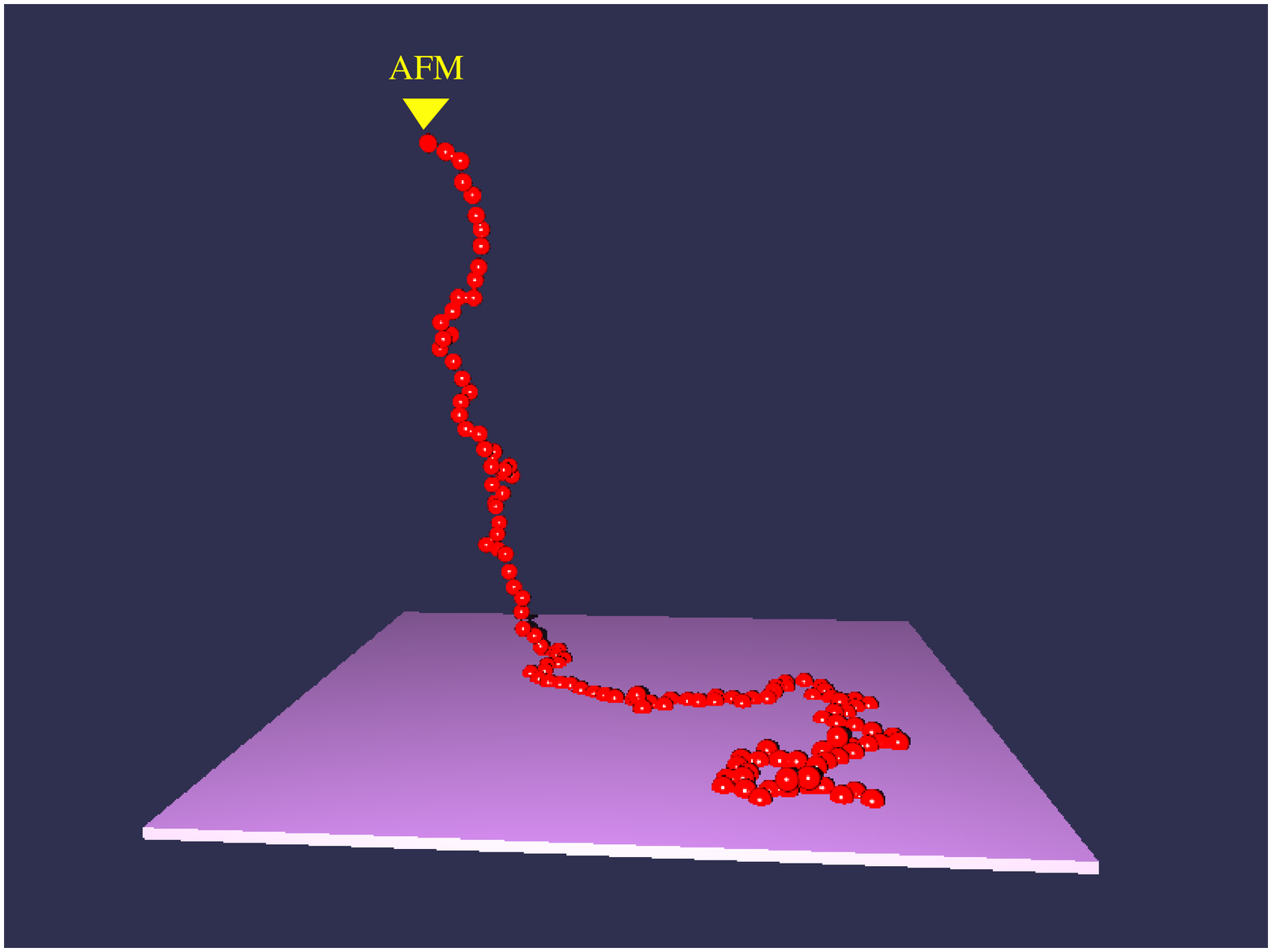}
\hspace{0.50cm}
\includegraphics[scale=0.55]{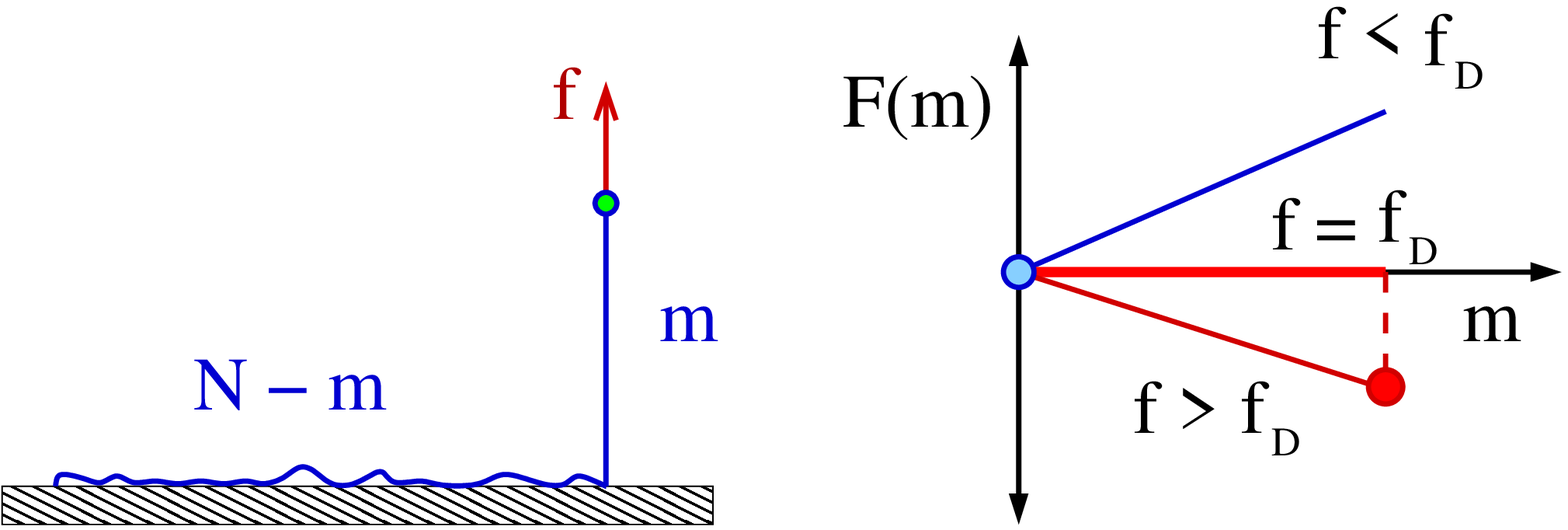}
\caption{(left) A snapshot from the MC simulation: $N=128,\;
h=25.0,\;\epsilon=4.0,; \langle f\rangle = 6.126$. (right) Schematic
representation of an adsorbed chain of length $N$ which is pulled by the end
segment off the surface with force $f$. While $N=m$ monomers lie on the plane,
$m$ monomers form the chain tail. The variation of $F(m)$ with $m \in [1,N]$ for
super- $f > f_D$ and subcritical $f < f_D$ forces indicates two different minima
(filled circles) of $F$ at $m=N$ and $m=0$ for $f > f_D$ and $f < f_D$,
respectively.}
\label{scheme_F}
\end{figure}

In order to illustrate the problem with chain detachment under pulling, we start
with a simple example, cf. Fig.~\ref{scheme_F}, which shows schematically a case
when $N - m$ chain monomers are adsorbed on the plane while the remaining $m$
monomers form a stretched tail subjected to external force $f$. Consider for
simplicity a phantom chain with no excluded volume interactions between the
segments. The partition function of such Gaussian chain can be written as
$\Omega (m) = \mu_2 ^{N-m} \exp[\epsilon (N-m) - mfa/k_BT] $ where $\mu_2$
denotes the so called connective constant in $d=2$ dimensions (e.g., $\mu_2 =
2.6$ on a cubic lattice). The dimensionless adsorption energy $\epsilon =
\varepsilon/k_BT$ measures the energy gain per contact with the surface while
the work to detach and move $m$ beads a distance $a$ away from the plane is
$af/k_BT$.

Evidently, the corresponding free energy $F/k_BT = -\ln \Omega(m) \approx - N
(\ln \mu_2 + \epsilon) + m[(\ln \mu_2 + \epsilon) - fa/k_BT]$ grows or declines
with varying $m$, depending on the sign of the expression in square brackets.
Therefore, one can readily define a critical detachment force $f_D(\epsilon) =
k_BT (\ln \mu_2 + \epsilon)/a$ such that for $f < f_D$ one finds a minimum of
$F$ at $m = 0$ (the chain is completely adsorbed) whereas for $f > f_D$ the
lowest free energy is reached for $m = N$ whereby the polymer is entirely
detached from the surface - Fig.~\ref{scheme_F}. At the critical value $f = f_D$
the free energy becomes independent of $m$, indicating even within this
oversimplified consideration (which neglects the presence of loops in the
adsorbed state) that {\em any} number of chain contacts with the adsorbing plane
becomes equally probable. Evidently, by just crossing the critical line
$f_D(\epsilon)$ the polymer chain undergoes an abrupt transition between an
adsorbed and detached state at any strength of adsorption $\epsilon$ whereby for
$f = f_D$ no states with a particular value of $m$ can be singled out as the
most probable. Physically this means that for $f = f_D$ one expects very strong
fluctuation of the number of contacs (which is our order parameter).
In the following we show that this simplified consideration is indeed confirmed
by the more general adsorption model too.

\subsection{The Grand Canonical Ensemble approach to chain adsorption}

\begin{figure}[htb]
\includegraphics[scale=0.5]{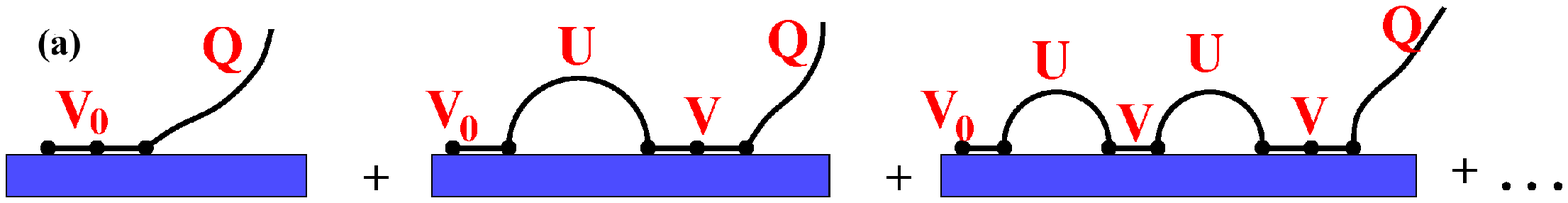}
\hspace{.50cm}
\includegraphics[scale=0.37]{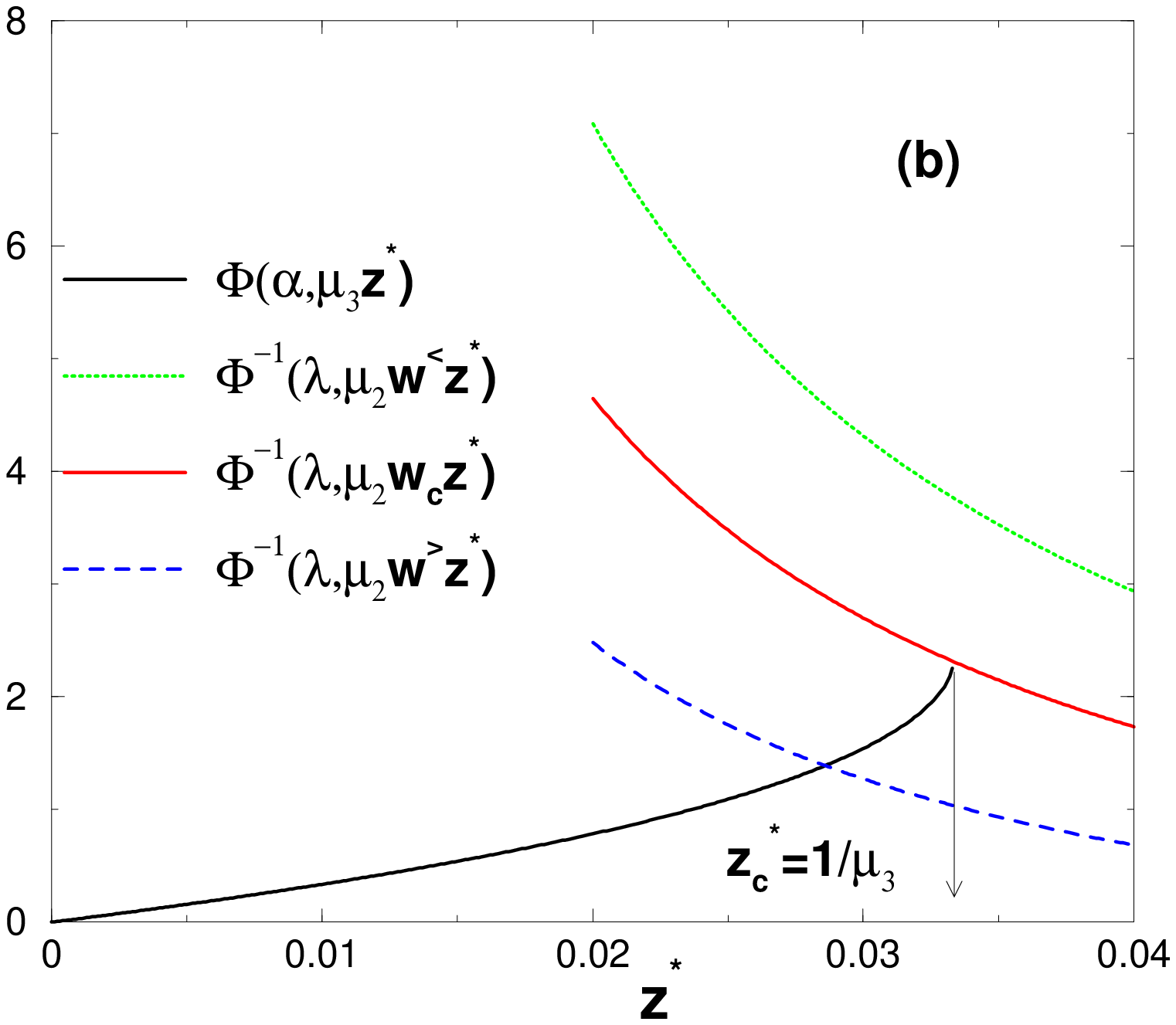}
\caption{(a) Schematic representation of the series expansion,
Eq.~(\ref{GC_partition}). (b) The intersection of the polylog functions
$\Phi(\alpha, \mu_3 z^*)$ and $1/\Phi(\lambda, \mu_2 w z^*)$ yields a solution
of Eq.~(\ref{Basic_Eq}) for the fugacity $z^*$. For adsorption strength
$\epsilon < \epsilon_c$ the corresponding Boltzmann weight $w^< =
\exp(\epsilon)$ is insufficient to provide an intersection point (the chain is
desorbed) whereas for $\epsilon > \epsilon_c$ (for $w^>$) a solution for $z^*$
exists (the chain is adsorbed). The CAP $\epsilon = \epsilon_c$ (i.e. for $w_c$)
is marked by the first appearance of common point of intersection (full lines)
at $z_c^* = 1/\mu_3$.}
\label{z_crit}
\end{figure}

Starting with the conventional (i.e., force-free) case of polymer adsorption, we
recall that an adsorbed chain is build up from loops, trains, and a free tail.
One can treat statistically these basic structural units by means of the GCE
approach\cite{Poland,Bir} where the lengths of the buildings blocks are not
fixed but may rather fluctuate. The GCE-partition function is then given by an
expansion over all possible lengths $N$, see Fig.~\ref{z_crit}a, which can be
considered and summed as a geometric series:
\begin{eqnarray}
 \Xi (z) = \sum_{N=0}^{\infty} \: \Xi_{N} \: z^{N} = \frac{V_{0}(z) \:
Q(z)}{1 - V(z) U (z)}.
\label{GC_partition}
\end{eqnarray}
In Eq.~(\ref{GC_partition}) $z$ is the fugacity and $U(z)$, $V(z)$, and $Q(z)$
denote the GCE partition functions of loops, trains and tails, respectively. The
building block adjacent to the tethered chain end is allowed for by $V_{0}(z) =
1+V(z)$. The partition function of the loops is defined as $ U (z)  =
\sum_{n=1}^{\infty} \: (\mu_3 z)^{n}/n^{\alpha}$, where $\mu_3$ is the $3d$
connective constant and $\alpha$ is the exponent which governs surface loops
statistics. It is well known that for an {\it isolated } loop $\alpha =
1-\gamma_{11} \approx 1.39$ \cite{Vanderzande} where $\gamma_{11}=- 0.390$. One
can prove\cite{SBVRAMTV} that $\alpha$ changes value, provided  the excluded
volume interactions between a loop and the rest of the chain are taken into
account. The train GCE-partition function reads $V(z)  = \sum_{n=1}^{\infty} \:
(\mu_3 w z)^{n}/n^{\lambda}$ with $1-\gamma_{d=2} \approx - 0.343$ whereby one
assumes that each adsorbed segment gains an additional statistical weight $w =
\exp(\epsilon)$. Eventually, the GCE partition function for the chain tail
is defined by $Q(z) = 1 + \sum_{n=1}^{\infty} \: (\mu_3 z)^{n} / n^{\beta}$. For
an isolated tail $\beta = 1-\gamma_1 \approx 0.32$ where $\gamma_1 =
0.680$\cite{Vanderzande} but again the excluded volume interactions of a tail
with the rest of the chain increase the value of $\beta$.

If one knows the GC partition function, Eq.~(\ref{GC_partition}), one can find
the number of weighted configurations of a polymer chain, containing $N$
segments (i.e., the canonical partition function of such chain), $\Xi_{N}$, by
taking the inverse Laplace transform of $\Xi(z)$. Using the {\em generating
function} method \cite{Rudnick}, one finds that the main contribution to the
coefficient $\Xi_N$ at $z^N$ is $(z^*)^{-(N+1)}$ which is provided by the
singularity at $z^*$ of $\Xi(z)$. There is a simple pole in
Eq.~(\ref{GC_partition}) at $z = z^*$, namely, when $V(z^*) U(z^*) = 1$. Thus 
one gets the free energy as $F=k_BTN \ln z^*$ and the fraction of adsorbed
monomers (which defines a convenient order parameter for the phase transition)
is $n = - \partial \ln z^*/\partial \ln w$. In terms of the so called {\it
polylog function}, which is defined as $\Phi (\alpha, z)=\sum_{n=1}^{\infty} \:
z^{n}/n^{\alpha}$ \cite{Erdelyi} and exists only for $z \le 1$, the equation for
$z^*$ reads
\begin{eqnarray}
\Phi(\alpha, \mu_3 z^*) \Phi(\lambda, \mu_2 w z^*) = 1.
\label{Basic_Eq}
\end{eqnarray}
A nontrivial solution for $z^*$ in terms of $w$ (or the adsorption energy
$\epsilon$) appears at the {\em critical adsorption point} (CAP) $w=w_c$  -
see Fig.~\ref{z_crit}b - where $\mu_3z^* = 1$. For example, close to the CAP
one may expand $\Phi(\alpha, \mu_3 z^*)$ with respect to $1-\mu_3z^*$ so
that $w_c$ is determined from $ \zeta (\alpha) \Phi(1-\gamma_{d=2}, \mu_2
w_c/\mu_3) = 1$ where  $\zeta(x)$ denotes the Riemann $\zeta$-function.
In the vicinity of the CAP the solution attains the form
\begin{eqnarray}
 z^*(w) \approx  [1 -  A \: (w - w_c)^{1/(\alpha-1)}]\mu_3^{-1}
\label{Solution}
\end{eqnarray}
where $A$ is a constant. Then, for the average fraction of adsorbed monomers one
obtains $n \propto (\epsilon - \epsilon_c)^{1/(\alpha-1) - 1}$. A comparison
with the well known scaling relationship $n \propto (\epsilon -
\epsilon_c)^{1/\phi - 1}$ where $\phi$ is the so called {\it adsorption (or,
crossover) exponent} \cite{Vanderzande} suggests that
\begin{eqnarray}
 \phi = \alpha - 1
\label{Phi}
\end{eqnarray}
This result, derived first by Birshtein\cite{Bir}, is of principal importance.
It shows that the exponent $\phi$, which describes polymer adsorption at
criticality, is determined by the value of $\alpha$ which governs the polymer
loop statistics! If loops are treated as isolated objects, then $\alpha =
1-\gamma_{11}\approx 1.39$ so that $\phi = 0.39$. In contrast, excluded volume
interactions between a loop and the rest of the chain lead to an increase of
$\alpha$ and $\phi$, as shown below.

From the expression for $U(z)$, given above, and Eq.~(\ref{Solution}) we have
$P_{\rm loop} \approx (\mu_3 z^*)^l/l^{1+\phi} \approx \exp[ - c_1 (\epsilon -
\epsilon_c)^{1/\phi}] / l^{1+\phi}$. This is valid only for $\epsilon >
\epsilon_c$ since a solution for Eq.~(\ref{Basic_Eq}) for subcritical values of
the adhesive potential $\epsilon$  does not exist. Nontheless, even in the
subcritical region, $\epsilon < \epsilon_c$, the monomers 
occasionally touch the substrate, creating thus single loops at the expense of
the tail length. The partition function of such a loop-tail configuration is
$Z_{l-t} = \frac{\mu_3^l}{l^{1+\phi}}\;\frac{\mu_3^{N-l}}{(N-l)^\beta}$. On the
other hand, the partition function of a tail conformation with no loops
whatsoever (i.e., of a nonadsorbed tethered chain) is $Z_t =
\mu_3^N \; N^{\gamma_1 - 1}$. Thus the probability $P^<_{\rm loop}(l)$ to find a
loop of length $l$ next to a tail of length $N-l$ can be estimated as $ P^<_{\rm
loop}(l) = \frac{Z_{l-t}}{Z_t} \propto \frac{N^{1 - \gamma_1}}{l^{1 + \phi}
(N - l)^{\beta}}$ for $\epsilon < \epsilon_c$. In the vicinity of the CAP,
$\epsilon \approx  \epsilon_c$, the distribution will be given by an
interpolation between the expressions above. Hence, the overall loop
distribution becomes
\begin{eqnarray}
P_{\rm loop}(l) =  \begin{cases}
                      \frac{1}{l^{1+\phi}}\exp\left [ -c_1 (\epsilon
-\epsilon_c)^{1/\phi}\;l\right ], \quad & \quad \epsilon >
\epsilon_c\\
\frac{A_1}{l^{1+\phi}} + \frac{A_2 N^{1-\gamma_1}}{l^{1+\phi} (N-l)^{\beta}},
\quad & \quad \epsilon = \epsilon_c\\
\frac{N^{1-\gamma_1}}{l^{1+\phi} (N-l)^{\beta}}. \quad & \quad  \epsilon <
\epsilon_c
                  \end{cases}
\label{Loop_distribution__all}
\end{eqnarray}
The same reasonings for a tail leads to the distribution
\begin{eqnarray}
P_{\rm tail}(l) =  \begin{cases}
                      \frac{1}{l^\beta}\exp\left [ -c_1 (\epsilon
-\epsilon_c)^{1/\phi}\;l\right ], \quad & \quad \epsilon >
\epsilon_c\\
\frac{B_1}{l^\beta} + \frac{B_2 N^{1-\gamma_1}}{l^\beta (N-l)^{1+\phi}}, \quad
& \quad \epsilon = \epsilon_c\\
\frac{N^{1-\gamma_1}}{l^\beta (N-l)^{1+\phi}}. \quad & \quad  \epsilon <
\epsilon_c
                  \end{cases}
\label{Tail_distribution__all}
\end{eqnarray}
In Eqs.~(\ref{Loop_distribution__all}) - (\ref{Tail_distribution__all}) $A_1,
A_2, B_1, B_2$ are constants. Close to the CAP these distributions are expected
to attain a U~-~shaped form (with two maxima at $l \approx 1$ and $l \approx
N$), as predicted for a Gaussian chain by Gorbunov {\it et al.} \cite{Gorbunov}.

For the average loop length $L$ the GCE-partition function for loops yields $L
= z \partial U(z)/\partial z |_{z=z^*} = \Phi (\alpha - 1, \mu_3 z^*)/\Phi
(\alpha, \mu_3 z^*)$. At the CAP, $L$ diverges as $L \propto 1/(\epsilon -
\epsilon_c)^{1/\phi - 1}$.

The average tail length $S$ is obtained as $S = z \partial Q(z)/\partial z
|_{z=z^*} = \Phi (\beta-1, \mu_3 z^*)/[1+ \Phi (\beta, \mu_3 z^*)]$. Again,
using the polylog function, one can show that at $\epsilon_c$  the average tail
length diverges as $S \propto 1/(\epsilon - \epsilon_c)^{1/\phi}$.

\subsection{The interaction of loops and the tail}

In the analytical expressions for the PDF of the different building
units of a chain,
Eqs.~(\ref{Loop_distribution__all})-(\ref{Tail_distribution__all}) we didn't
elaborate on the numerical values of the exponents $\alpha$ (that is, $\phi$)
 and $\beta$, taking as an example those for non-interacting polymer chains.
However, for a realistic self-avoiding chain one has to allow for the existence
of excluded-volume interactions. To this end one may consider the number of
configurations of a tethered chain in the vicinity of the CAP as an array of
loops which end up with a tail. Using the approach of Kafri {\it et al.}
\cite{Kafri} along with Duplantier's \cite{Duplantier} graph theory of polymer
networks, one may write the partition function $Z$ for a chain with ${\cal N}$
building blocks: ${\cal N}-1$ loops and a tail\cite{SBVRAMTV}. Consider now a
single loop of length $M$ while the length of the rest of the chain is $K$, that
is, $M+K=N$. In the limit of $M \gg 1,\; K\gg 1$ (but with $M/K \ll 1$) one can
show \cite{SBVRAMTV} that $Z \sim \mu_{3}^{M} \: M^{\gamma_{\cal N}^s -
\gamma_{{\cal N}-1}^s} \:\: \mu_{3}^{K} \: K^{\gamma_{{\cal N}-1}^s - 1}$ where
the surface exponent $ \gamma_{\cal N}^s = 2  + {\cal N}(\nu + 1) + \sigma_1 +
\sigma_1^s$ and $\sigma_1,\; \sigma_1^{s}$ are critical bulk and surface
exponents \cite{Duplantier}. The last result indicates that the effective loop
exponent $\alpha$ becomes
\begin{equation}
 \alpha = \gamma_{{\cal N}-1}^s - \gamma_{\cal N}^s = \nu + 1
\label{Effective_alpha}
\end{equation}
Thus, $\phi=\alpha-1=\nu = 0.588$, in agreement with earlier Monte Carlo
findings \cite{Eisenriegler}. One should emphasize, however, that the foregoing
derivation is Mean-Field-like ($Z$ appears as a product of loop- and
rest-of-the-chain contributions) which overestimates the interactions and
increases significantly the value of $\alpha$, serving thus as an upper bound
estimate. The value of $\alpha$, therefore, is found to satisfy the inequality
$1-\gamma_{11}\le \alpha \le 1+\nu$, i.e., depending on loop interactions, $0.39
\le \phi \le 0.59$.

\subsection{Taking the pulling force into account}

The GCE approach, described above, can now be employed to tackle the case of
self-avoiding polymer chain adsorption in the presence of pulling force. Thus 
we extend the consideration of Gaussian chains by Gorbunov {\it et al.}
\cite{Skvortsov}.

As far as a force $f$ is applied to the end-monomer of a tethered chain, one may
choose two possible ways in which the chain detachment from the adsorbing
surface can be carried out. One may fix $f$ as an independent control parameter
and study the variation of the height $h$ of the end-monomer above the surface
plane which corresponds to treatment within the constant force ensemble,
herafter referred to as $f$-ensemble. Or, one might fix $h$ and measure the
force acting on the end-monomer at a given height, working thus in the constant
height ensemble which we call in what follows the $h$-ensemble.

\subsubsection{$f$-ensemble}

\begin{figure}[htb]
\includegraphics[scale=0.44]{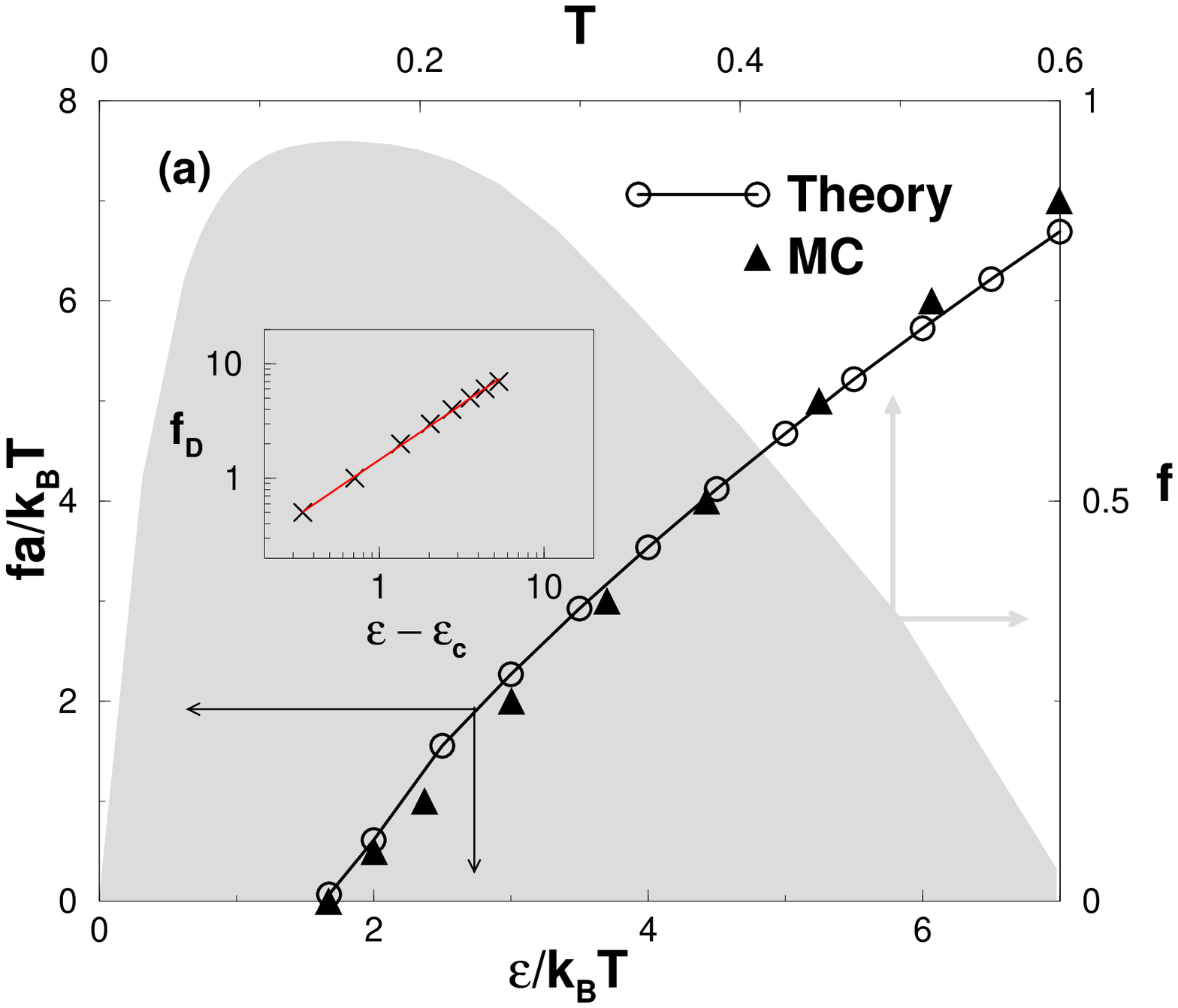}
\hspace{0.50cm}
\includegraphics[scale=0.44]{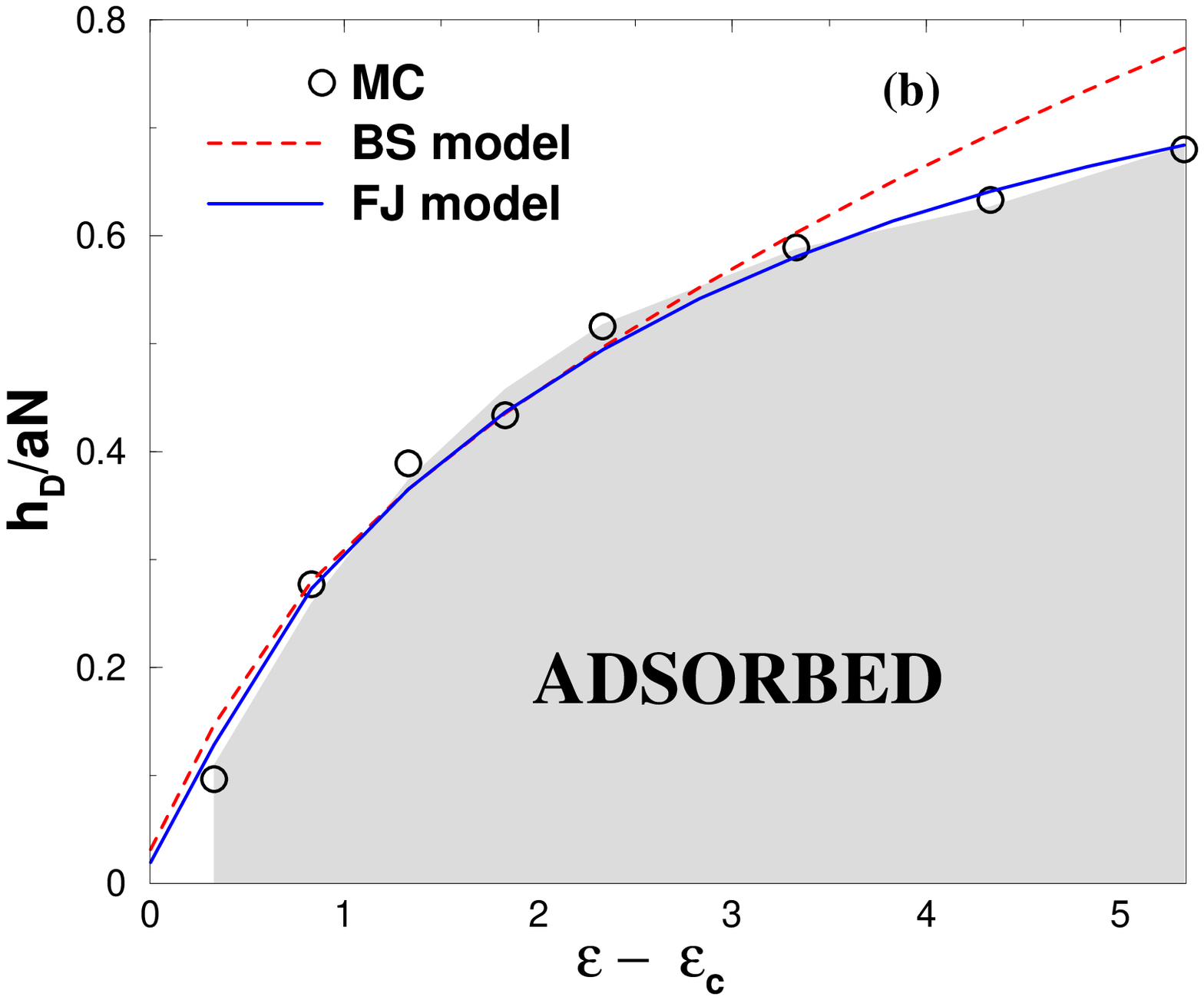}
\caption{ (a) Plot of the critical detachment force $f_D=fa/k_BT$ against the
surface potential $\varepsilon/k_BT$. Full and empty symbols denote MC and
theoretical results. A double logarithmic plot of $f_D$ against $\epsilon -
\epsilon_c$ with $\epsilon_c = 1.67$ is shown in the inset, yielding a slope of
$0.97\pm 0.02$, in agreement with the prediction $f_D \propto
(\epsilon-\epsilon_c)^{\nu/\phi}$. Shaded is shown the same phase diagram,
derived by numeric solution of Eq.~(\ref{Basic_Eq}) along with $z^*(w) =
z^\#(\tilde{f})$, which in dimensional $f$ (right axis) against $T$ (top axis)
units appears {\em reentrant}. (b) The same phase diagram in units of
detachment height $h_D$ and the distance from the CAP $\epsilon - \epsilon_c$.
Dashed and solid lines denote theoretical predictions based on the Pincus, or
Langevin force vs. elongation relationship while symbols show simulation data.}
\label{Phase-diagram}
\end{figure}

Under pulling force $f$, the tail GCE-partition
function $Q(z)$ in Eq.~(\ref{GC_partition}) has to be replaced by $\tilde{Q}(z)
= 1 + \sum_{n=1}^{\infty} [(\mu_3 z)^n / n^{\beta}] \: \int d^3 r P_n({\bf r})
\exp(f r_{\perp}/T)$ where $P_n({\bf r})$ is the end-to-end distance
distribution function for a self-avoiding chain \cite{DesCloizeaux} and $f
r_{\perp}$ measures the work, spent to pull the chain end to height $r_{\perp}$
above the adsorbing surface. After some
straightforward calculations $\tilde{Q}(z)$ can be written as
\begin{eqnarray}
\tilde{Q}(z) = 1 + a_1 \:{\tilde f}^{1-\gamma_1/\nu} \: \Phi (1-\nu, z \mu_3
\exp (a_2
{\tilde f}^{1/\nu})
\label{GC_Tail_Force}
\end{eqnarray}
with the dimensionless force ${\tilde f}=fa/k_BT$. The function $\tilde{Q}(z)$
has a {\em branch point} singularity at $z^{\#}=\mu_3^{-1} \exp(-a_2 {\tilde
f}^{1/\nu})$, i.e., $\tilde{Q}(z) \sim 1/(z^{\#}-z)^{\nu}$. One may,
therefore, conclude that the total GCE-partition function $\Xi(z)$ has two
singularities on the real axis: the pole $z^*$, related to the CAP, and the
branch point $z^{\#}$, related to the pulling force. It is known (see, e.g.,
Sec. 2.4.3. in \cite{Rudnick}) that for $N \gg 1$ the main contributions to
$\Xi_N$ come from the pole and the branch singular points, i.e.,
\begin{eqnarray}
 \Xi_N \sim C_1 \: (z^{*})^{-N} + \frac{C_2}{\Gamma(\nu)} \: N^{\nu-1} \:
(z^{\#})^{-N}
\label{Pole_Branch}
\end{eqnarray}
Evidently, for large $N$ only the smallest of these points matters. Note that
$z^{*}$ depends on the adsorption energy $\epsilon$ only (through
$w=\exp(\epsilon)$) whereas $z^{\#}$ is controlled by the external force
${\tilde f}$. Therefore, in terms of the two {\it control parameters},
$\epsilon$ and ${\tilde f}$, the equation $ z^{*}(\epsilon_D) = z^{\#}({\tilde
f}_D)$ defines the critical transition line between the adsorbed phase and the
force-induced desorbed phase - Fig.~\ref{Phase-diagram}. In the following
this line will be called {\it detachment line} (DL). Below it, $f < f_D$, or
above, $f > f_D$, either $z^*$ or $z^\#$, respectively, contribute to $\Xi_N$.
The controll parameters, $\epsilon_D$ and ${\tilde f}_D$, which satisfy  this
equation, denote detachment energy and detachment force, respectively.

On the DL the system undergoes a {\it
first-order phase transition}. The DL itself ends for ${\tilde f}_D
\rightarrow 0$ in the CAP, $\epsilon_c$, where the transition becomes of {\em
second} order, as is known for polymer adsorption without pulling. In the
vicinity of the CAP the detachment force ${\tilde f}_D$ is predicted to vanish
as ${\tilde f}_D \sim (\epsilon - \epsilon_c)^{ \nu / \phi}$.

This first order adsorption-desorption phase transition under pulling has a
clear {\em dichotomic} nature (i.e., it follows an ``either - or'' scenario): in
the thermodynamic limit $N \rightarrow \infty$ there is {\em no phase
coexistence}! The configurations are divided into adsorbed and desorbed
dichotomic classes. Metastable states are completely absent. Moreover, the mean
loop length $L$ remains {\em finite} upon DL crossing.  In contrast, the average
tail length $S$ {\em diverges} close to the DL. Indeed, at ${\tilde f} < {\tilde
f}_D$ the average tail length is given by $S={\tilde f}^{1 - \gamma_1/\nu}
\Phi (-\nu, z^*(w)/z^{\#}({\tilde f}))/[1+a_1 \Phi (1-\nu,
z^*(w)/z^{\#}({\tilde f}))]$. At the DL, $z^*=z^{\#}$, it diverges as $S \propto
{\tilde f}_D/({\tilde f}_D - {\tilde f})$.

\subsubsection{$h$-ensemble}

In the constant height ensemble the way a chain tethered to a surface responds
to stretching is described by the tail partition function $\tilde{Q}_N$. The
partition function of such chain with a fixed distance $h$ of the chain end
from
the anchoring plane is given
\begin{eqnarray}
 \tilde{Q}_{\rm tail}(N, h) = \frac{\mu_{3}^{N}}{N^{\beta}} \: a P_{N} (h)
\label{Partition_Tail}
\end{eqnarray}
where again $\beta=1-\gamma_1$ and $a$ is the bond length. The deformation of a
polymer chain can be described within two models: the bead-spring (BS) model
for flexible bonds and the freely jointed chain (FJ) model in which the bonds
between monomers are considered rigid. In the BS model one can use for $P_N$ the
expression \cite{Kreer}
\begin{eqnarray}
 P_N(h)  = \frac{A}{R_N} \: \left(\frac{h}{R_N}\right)^{\zeta} \: \exp \left[-D
\left(\frac{h}{R_N}\right)^{1/(1-\nu)}\right]
\label{Overall}
\end{eqnarray}
where the exponent $\zeta \approx 0.8$, and $A$ is a normalization constant. The
free energy of the tethered chain with a fixed distance $h$ takes on the form
$F_{\rm tail}(N, h) = - T \ln \tilde{Q}_{\rm tail} (N, h)$.  By making use of
Eqs. (\ref{Partition_Tail}) and (\ref{Overall}), the expression for the force
$f_N$, acting on the end-monomer when kept at distance $h$ is given by
\begin{eqnarray}
 f_N = \frac{\partial}{\partial h} \: F_{\rm tail}(N, h) 
= \frac{k_BT}{R_N} \:\left[ \frac{D}{1-\nu} \left(
\frac{h}{R_N}\right)^{\nu/(1-\nu)} - \zeta \left( \frac{R_N}{h}\right)\right]
\label{Deformation}
\end{eqnarray}
One should note that at $h/R_N \gg 1$ we recover the well known Pincus
deformation law: $h \propto a N (a f_N/k_BT)^{1/\nu -1}$. In this approximation
the (dimensionless) elastic energy reads $U_{\rm el}/k_BT = - N (a
f_{N}/k_BT)^{1/\nu} $. In result the corresponding tail free energy is given
by \begin{eqnarray} \frac{F_{\rm tail}}{k_BT} =   - N \left(\frac{a
f_N}{T}\right)^{1/\nu} - N \ln \mu_3 \label{Free_Energy_BS}
\end{eqnarray}
Eq.~(\ref{Deformation}) indicates that there exists a height $h_0 = (\zeta
(1-\nu)/ D)^{1-\nu} R_N$ over the surface where the force $f_N$ changes
sign and becomes negative (that is, the surface repulsion dominates). According
to Eq.~(\ref{Deformation}) the force diverges as $f_N \propto - k_BT/h$ upon
further decrease of the distance $h$.

It is well known \cite{Grosberg} that the Pincus law, Eq. (\ref{Deformation}),
describes the deformation behavior at intermediate force strength, $1/N^{\nu}
\ll a f_N/k_BT \leq 1$. Direct Monte Carlo simulation results indicate that,
depending on the model, deviations from Pincus law emerge at $h/R_N \ge 3$
(bead-spring off-lattice model) \cite{Lai}, or $h/R_N \ge 6$ (Bond
Fluctuation Model) \cite{Wittkop}. In such ``overstretched'' regime (when the
chain is stretched close to its contour length) one should take into account
that
the chain bonds cannot expand indefinitely. This case could be treated within
the simple FJ model \cite{Lai} where the bond length $a$ is fixed. In this model
the force - elongation relationship is given by
\begin{eqnarray}
 f_N = \frac{k_BT}{a} \: {\cal L}^{-1} \left( \frac{h}{a N}\right)
\label{L}
\end{eqnarray}
where ${\cal L}^{-1}$ denotes the inverse Langevin function ${\cal
L}(x) = \coth(x) - 1/x$. The corresponding free energy of the tail for the FJ
model reads
\begin{eqnarray}
 \frac{F_{\rm tail}}{k_BT} =   - N {\cal G}\left(\frac{a f_N}{T} \right)  - N
\ln \mu_3
\label{Free_Energy_FJBV} 
\end{eqnarray}
where we have used the notation ${\cal G}(x) = x {\cal L}(x) = x \coth (x) - 1$.
One should emphasize that the force $f_{N}$ stays constant in the course of the
pulling process (i.e., as long as one monomer, at least, is adsorbed on the
surface), thus  $f_{N}$ corresponds to a {\it plateau} on the elongation curve
$f - h$. An adsorbed monomer has a chemical potential, $\mu_{\rm ads}=\ln
z^{*}$, which should be equal in equilibrium to the chemical potential of a
desorbed monomer in the tail, $\mu_{\rm des}= \partial (F_{\rm tail}/T)/\partial
N$.  Thus the condition $\mu_{\rm
ads}=\mu_{\rm des}$ leads to the following ``plateau law'' relationship
\begin{eqnarray}
 \frac{a \: f_{\rm p}}{k_BT} = \begin{cases}\left|\ln [\mu_3
z^{*}(\epsilon)]\right|^{\nu} &\mbox{,  BS model}\\
\\
{\cal G}^{-1}\left(\left|\ln [\mu_3 z^{*} (\varepsilon)]\right|\right) &\mbox{,
  FJ model}
\end{cases}
\label{Local_Equilibrium}
\end{eqnarray}
with ${\cal G}^{-1}$ being the inverse of the ${\cal G}$ function.  Close to the
critical point $\epsilon_c$ the plateau force $f_{\rm p}\to 0$. Indeed, taking
into account that in the vicinity of the critical point $\ln [\mu_3
z^{*}(\epsilon)] \propto - (\epsilon - \epsilon_c)^{1/\phi}$ \cite{SBVRAMTV} and
${\cal G}^{-1} (x)\approx (3 x)^{1/2}$ we conclude that $f_{\rm p} \propto
(\epsilon - \epsilon_c)^{\nu/\phi}$ for the BS model and $f_{\rm p} \propto
(\epsilon - \epsilon_c)^{1/2 \phi}$ for the FJ model. If the number of
tail monomers is denoted by $M$, then the  one can write
\cite{SBVRAMTV}  $n = - (1/T N) \partial F_{\rm ads}/\partial
\epsilon $, where $F_{\rm ads}$ is  the free energy of the adsorbed portion of
the chain given as  $F_{\rm ads} = k_BT [ N - M(h, \epsilon)] \ln
z^{*}(\epsilon)$. From Eq.~(\ref{Local_Equilibrium}) one can easily obtain $M$
for $h\gg R_g$ so that in result one gets
\begin{eqnarray}
 n &=&  - \left[1 - \frac{M(h, \epsilon)}{N}\right] \frac{\partial \ln
z^{*}(\epsilon)}{\partial
\epsilon} + \frac{\ln z^{*}(\epsilon)}{N} \: \frac{\partial  M(h,
\epsilon)}{\partial \epsilon} = \nonumber \\
&=& \left|\frac{\partial \ln z^{*}(\epsilon)}{\partial
\epsilon}\right| \times \begin{cases}
                         1 - \frac{h}{c_2 a N} \left(\frac{k_BT}{a f_{\rm
p}}\right)^{1/\nu - 1} \left[1 - {\tilde c}_1 \left(\frac{k_BT}{a f_{\rm
p}}\right)^{1/\nu}| \ln z^{*}|\right] &\mbox{, for BS-model}\\
\\
1 - \frac{h}{c_3 a N} \left[{\cal L} \left(\frac{a f_{\rm
p}}{k_BT}\right)\right]^{-1} \left[1 - \frac{{\cal L}'\left(\frac{a f_{\rm
p}}{k_BT}\right)}{{\cal L}\left(\frac{a f_{\rm p}}{k_BT}\right) {\cal
G}'\left(\frac{a f_{\rm p}}{k_BT}\right)} \: c_1 \: |\ln
z^{*}|\right] &\mbox{, for FJ-model}
\end{cases}
\label{Order_Parameter}
\end{eqnarray}
where $c_1,\;c_2,\;c_3$ are constants of the order of unity. The derivatives
${\cal L}'(x) = 1/x^2 - 1/[\sinh(x)]^2 $ and  ${\cal G}'(x) = {\cal L}(x) + x
{\cal L}'(x)= \coth(x) - x /[\sinh (x)]^2$ and ${\tilde c}_1 = (1 - \nu) c_1$.

As one can see from Eq.~(\ref{Order_Parameter}), the order parameter
{\em decreases linearly} and steadily (no jump!) with growing $h/N$.

\subsection{Reentrant phase behavior}

Recently, it has been realized \cite{Mishra} that the DL, force $f_D$ versus
temperature $T$, when represented in units with dimension, goes (at a relatively
low temperature) through a maximum, i.e., the desorption transition shows
reentrant behavior! Such behavior has been predicted
earlier\cite{Orlandini,Maren,Kumar1} in a different context, namely, of
DNA-unzipping, and also in the coil-hairpin transition\cite{Kumar2}.

One can readily see that this result follows directly from our theory. Indeed,
the solution of Eq.~(\ref{Basic_Eq}) at large values of $\epsilon$ (that is, at
low temperature) can be written as $z^* \approx {\rm e}^{- \epsilon} / \mu_3$ so
that the DL, $z^{*}=z^{\#}$, in terms of {\it dimensionless} parameters is
monotonous, ${\tilde f}_D  \propto [\epsilon_D - \ln(\mu_3 / \mu_2)]^{\nu}$.
Note, however, that the same DL, if represented in terms of the {\it
dimensional} control parameters, force $f_D$ versus temperature $T_D$ (with a
fixed energy $\varepsilon_0$), shows a nonmonotonic behavior $f_D = k_B T_D
[\varepsilon_0/T_D  - \ln(\mu_3/\mu_2)]^{\nu}/a$ - Fig.~\ref{Phase-diagram}, as
found earlier for DNA-unzipping \cite{Orlandini}. This curve has a maximum at a
temperature given by $k_B T_{D}^{max} = (1-\nu)\varepsilon_0/\ln(\mu_3/\mu_2)$.
At very low $T$, however, the expression for $P_n({\bf r})$ \cite{DesCloizeaux}
predicts divergent chain deformation \cite{Orlandini}, i.e., it becomes
unphysical. One can readily show that in this case the correct behavior is given
by $f a = \varepsilon_0 + k_BT\ln(\mu_3/\mu_2)$.

\section{Monte Carlo Simulation Model}\label{MC_model}

We use a coarse grained off-lattice bead-spring model \cite{MC_Milchev} which
has proved rather efficient in a number of polymers studies so far. The system
consists of a single polymer chain tethered at one end to a flat impenetrable
structureless surface - Fig.~\ref{scheme_F}. The surface interaction is
described by a square well potential,
 \begin{equation}\label{ads_pot}
 U_w(z) = \begin{cases}
\epsilon, & z < r_c \\
0, & z \ge r_c
          \end{cases}
\end{equation}
The strength $\epsilon$ is varied from $1.0$ to $7.0$ while the
interaction range $r_c = 0.125$. The effective bonded interaction is described
by the FENE (finitely extensible nonlinear elastic) potential:
\begin{equation}
U_{FENE}= -K(1-a)^2ln\left[1-\left(\frac{l-a}{l_{max}-a} \right)^2 \right]
\label{fene}
\end{equation}
with $K=20, l_{max}=1, a =0.7, l_{min} =0.4$. The nonbonded interactions
between monomers are described by the Morse potential:
\begin{equation}\label{Morse}
\frac{U_M(r)}{\epsilon_M} =\exp(-2\alpha(r-r_{min}))-2\exp(-\alpha(r-r_{min}))
\end{equation}
with $\alpha =24,\; r_{min}=0.8,\; \epsilon_M/k_BT=1$. In few cases, needed to
clarify the nature of the polymer chain resistance to stretching, we have taken
the nonbonded interactions between monomers as purely repulsive by shifting the
Morse potential upward by $\epsilon_M$ and removing its attractive branch,
$V_M(r) = 0$ for $r \ge r_{min}$.

We employ periodic boundary conditions in the $x-y$ directions and impenetrable
walls in the $z$ direction. The lengths of the studied polymer chains are
typically $64$, and  $128$. The size of the simulation box was chosen
appropriately to the chain length, so for example, for a chain length of $128$,
the box size was $256 \times 256 \times 256$ . All simulation results have been 
averaged over about $2000$ measurements.

The standard Metropolis algorithm was employed to govern the moves with  self
avoidance automatically incorporated in the potentials. In each Monte Carlo
update, a monomer was chosen at random and a random displacement attempted with
$\Delta x,\;\Delta y,\;\Delta z$ chosen uniformly from the interval $-0.5\le
\Delta x,\Delta y,\Delta z\le 0.5$. If the last monomer was displaced in $z$
direction, there was an energy cost of $-f\Delta z$ due to the pulling force.
The transition probability for the attempted move was calculated from the change
$\Delta U$ of the potential energies before and after the move was performed as
$W=exp(-\Delta U/k_BT)$. As in a standard Metropolis algorithm, the attempted
move was accepted, if $W$ exceeds a random number uniformly distributed in the
interval $[0,1]$.

As a rule, the polymer chains have been originally equilibrated in the MC method
for a period of about $ 5 \times 10^5$ MCS after which typically $500$
measurement runs were performed, each of length $2 \times 10^6$ MCS. The
equilibration period and the length of the run were chosen according to the
chain length and the values provided here are for the longest chain length.

\section{Comparison of Simulation Data with Theoretical
Predictions}\label{Results}

We have investigated the force induced desorption of a polymer performing MC
simulations in the $f$-ensemble and in the $h$-ensemble. As an order parameter
for the desorption transition we use the fraction of monomers $n$ in contact
with the sticky surface. Below we present few typical quantities of interest
which manifest the good agreement between theoretical predictions and simulation
results. Another important point is the observed qualitative difference between
the $f-$ and $h-$ensembles in the behavior of some basic properties like the
order parameter of the phase transition.
\vspace{0.70cm}
\begin{figure}[htb]
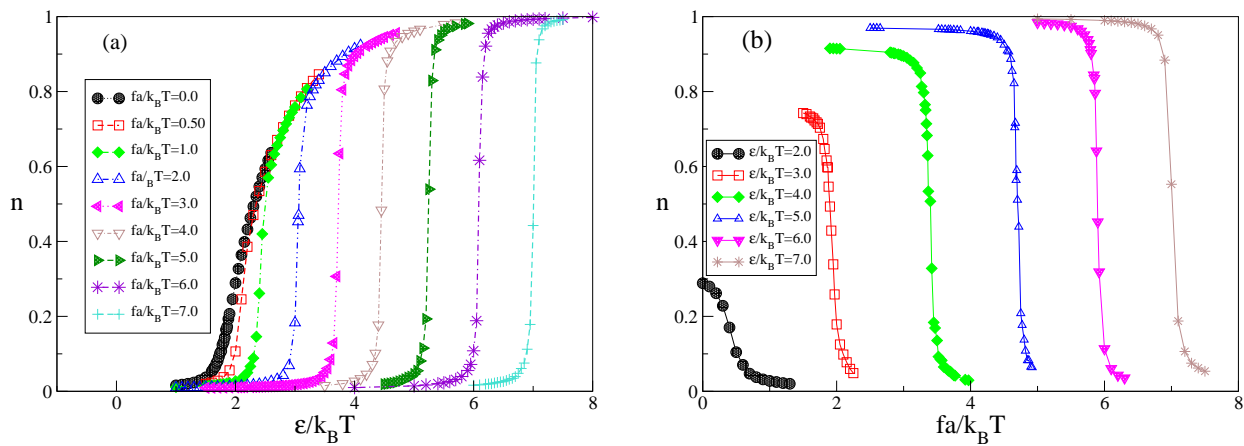

\includegraphics[scale=0.33]{n_e.eps}
\hspace{0.50cm}
\includegraphics[scale=0.33]{n_f.eps}
\caption{(a) The 'order parameter', $n$, against the surface potential,
$\epsilon$, for various pulling forces $f$. The chain has length $N$=128. (b)
Plot of $n$ vs. $f$ for several surface potentials $\epsilon$.}
\label{Order-Parameter}
\end{figure} 
Fig.~\ref{Order-Parameter}a shows the variation of the order parameter $n$ with
changing adhesive potential $\epsilon$ in the $f$-ensemble at fixed pulling
force whereas Fig.\ref{Order-Parameter}b depicts $n$ vs. force $f a/T$ for
various $\epsilon$. The abrupt change of the order parameter is in close
agreement with our theoretical prediction. Indeed, from
Fig.~\ref{Order-Parameter} one can readily verify that the polymer detachment
transition is of first order.

However, the order parameter variation in the equivalent $h$-ensemble looks
very different.
\vspace{0.50cm}
\begin{figure}[bht]
\hspace{-2.0cm}
\includegraphics[scale=0.35, angle=0]{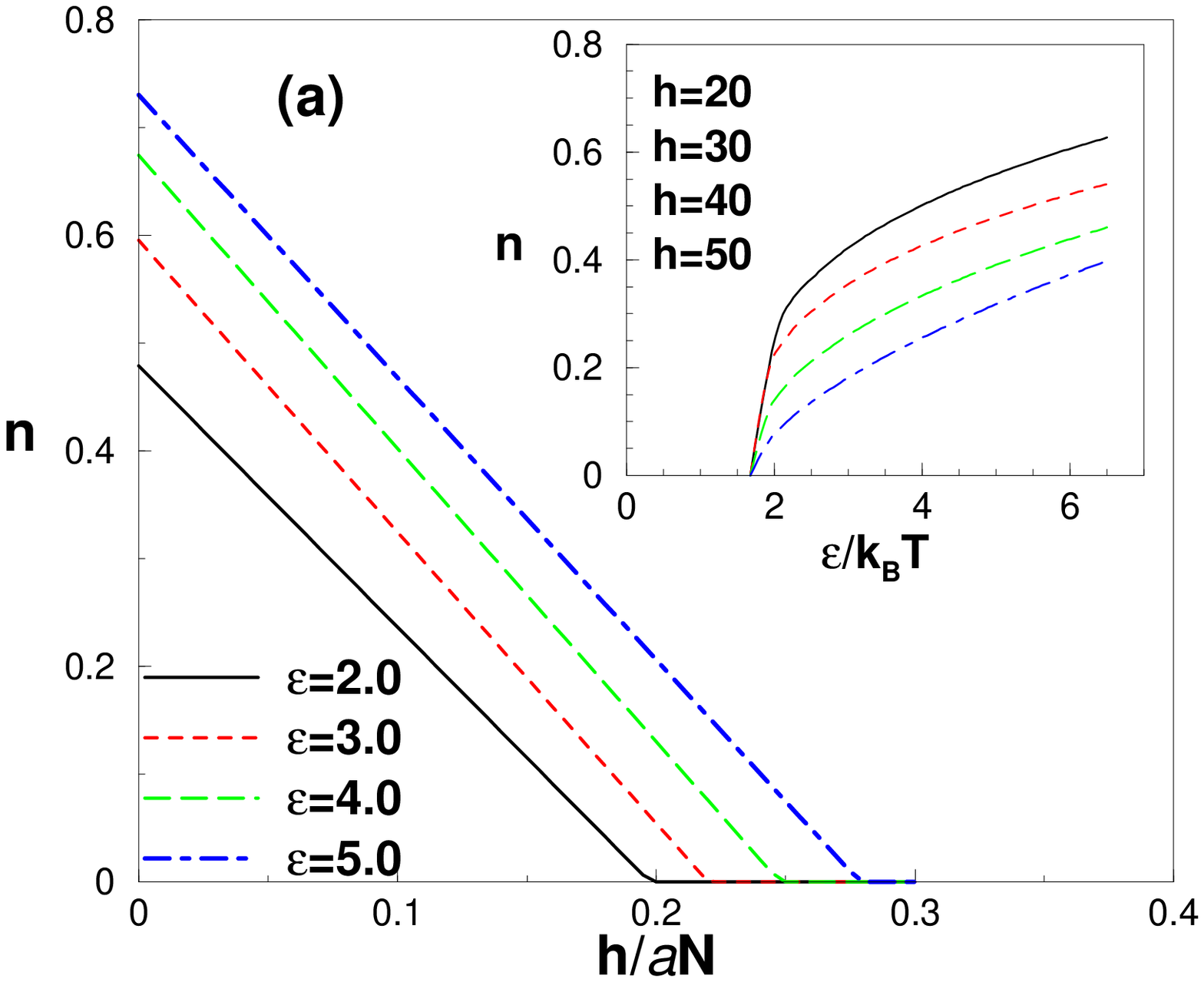}
\hspace{0.50cm}
\includegraphics[scale=0.29, angle=0]{op_MC.eps}
\caption{\label{OP} (a) Order parameter $n$ variation with changing height
$h/aN$ of the fixed chain-end for polymers of length $N = 64,\; 128$ and
different adsorption strength $\epsilon/k_BT$. (b) Variation of $n$ with
$\epsilon/k_BT$ for different fixed positions of the chain-end $h/aN$.}
\end{figure}
In Fig.~\ref{OP}a, \ref{OP}b, we show the change in $n$ with $h$ and in the
insets the variation of the fraction of adsorbed segments with adsorption
strengths $\epsilon$ for several fixed heights $20 \le h \le 50$ of the $N =
128$ chain. It is evident that, apart from the rounding of the MC data for $n$
at $n \to 0$, which is less pronounced for $N=128$ than for $N=64$, one finds
very good agreement between the behavior, predicted by
Eq.~(\ref{Order_Parameter}), and the simulation results. Comparing
Figs.~\ref{Order-Parameter} and \ref{OP} one realizes the striking difference
between the order parameter behavior in the $f-$ and $h-$ensembles. However, if
the height $h$ on the $x$-axis of Fig.~\ref{OP}a is expressed in terms of the
corresponding average force $\langle f \rangle$, one recovers again a jump in
the order parameter $n$ \cite{SKB}. 

The peculiar nature of the desorption transition under pulling becomes more
evident when one plots the PDF of the order parameter in both statistical
ensembles. In the presence of a pulling force one observes a remarkable feature
of the order parameter probability distribution - Fig.~\ref{fig_PDF_n}a:
\vspace{1.2cm}
\begin{figure}[bht]
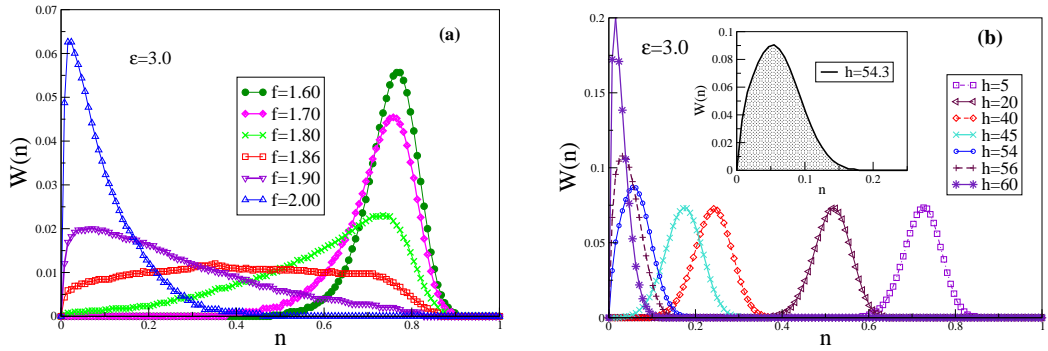

\includegraphics[scale=0.27]{ophistogram_f.eps}
\hspace{0.5cm}
\includegraphics[scale=0.27]{ophistogram_h.eps}
\vspace{0.5cm}
\caption{Order parameter distribution of a polymer with $N=128$ and ads
adsorption potential $\epsilon = 3.0$: (a) in the $f-$ensemble for several
values of $f$. The critical detachment force is $f_D \approx 1.85 \pm 0.01$. (b)
in the $h-$ensemble for several fixed heights $h$. The inset displays
the distribution $W(n)$ at the critical height of detachment $h_D \approx
54.3$.}
\label{fig_PDF_n}
\end{figure}
- an absence of two peaks in the vicinity of the transition force $f_D$
although bimodality is customary in first-order phase transition. Immediately
at $f_D$ the distribution $W(n)$ is {\em flat}, indicating huge fluctuations
of $n$ so that {\em any} value of the number of contacts is equally probable. 
This lack of bimodality in the $W(n)$ manifests the dichotomic nature of the
desorption transition which rules out phase coexistence. In contrast, in the
$h-$ensemble, Fig.~\ref{fig_PDF_n}b, one observes an entirely different 
shape of $W(n)$ with only slight deviations (an appearance of
non-zero third moment of the distribution) from Gaussianity in the vicinity of
$h_D$. The fluctuations of $n$, according to the half-width of $W(n)$, remain
finite and almost unchanged for all values of $h$.

\begin{figure}[bht]
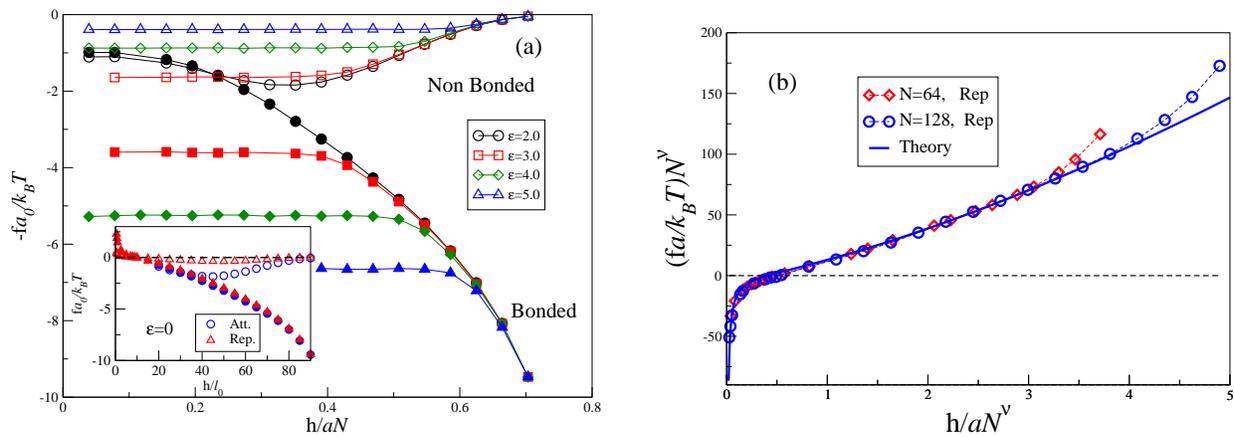

\includegraphics[scale=0.31]{f_h_plateau.eps}
\hspace{0.5cm}
\includegraphics[scale=0.31]{h_f_Pincus.eps}
\vspace{0.5cm}
\caption{ (a) Variation of the two components to the total force, exerted by the
chain on the end-monomer which is fixed at (dimensionless) height $h/aN$ for
different adsorption potentials $2.0 \le \epsilon/k_BT \le 5.0$ and  bonding
(FENE) interactions (full symbols) as well as non-bonding (Morse) interactions
(empty symbols). In the inset the same is shown for a neutral plane $\epsilon =
0.0$ and purely repulsive monomers (triangles) and for the usual Morse potential
(circles). (b) Variation of the total applied force $f$ with growing height of
the end monomer in terms of Pincus reduced variables, $faN^{\nu}/k_B T$ versus 
$h/aN^\nu$,  for a polymer with purely repulsive nonbonded forces for
$N=64,\;128$.}
\label{fig_f_h}
\end{figure}
Eventually, we show in Fig.~\ref{fig_f_h}a the typical plateau observed in the
average pulling force when the polymer detachment is effected in the
$h$-ensemble. Within a large interval of height variation the mean force,
exerted on the end monomer, remains constant as observed in laboratory
experiments. A rapid growth in the magnitude of this force sets in after the
plateau, as soon as the bonds rather that the confortmation of the polymer are
stretched upon further elongation. The stronger the adsorption, $\epsilon$, the
larger the force $f_D$ required to remove the chain from the substrate.

In addition to the force due to bonded interactions, however, one can see a
small contribution from the non-bonded (attractive) interactions between the
chain segments. This contribution is not allowed for by the GCE theory and,
therefore, a test with the theoretical preedictions should exclude it. If the
attractive branch of the Morse potential is removed, leaving the self-excluded
repulsive branch only, this contribution almost vanishes - Fig.~\ref{fig_f_h}a
(inset).

The elongation vs. force relationship, predicted by Eq.~(\ref{Deformation}), is
tested in Fig.~\ref{fig_f_h}b for chains in which only non-bonded repulsion
between segments exists. For small and intermediate extensions $h$ where $f$ is
not too large the agreement with Pincus law is found to be perfect although it
deteriorates for larger $f$, as expected. In the latter region one may show that
a very good agreement between theory and computer experiment is provided by the
FJ model - Eq.~(\ref{L}). From Fig.~\ref{fig_f_h}b one can also see that $f$
goes through zero {\em before} the height has become zero, that is, no force is
felt when the chain end is kept at this particular height. Further decrease of
$h$ leads to change of sign of $f$, indicating the entropic repulsion of the
polymer coil from the solid surface.

\section{Summary}

In conclusion, we have shown that a full description of the force-induced
desorption of a self-avoiding  polymer chain can be achieved by means
of the GCE approach, yielding the average size and probability distribution
functions of all basic structural units of partially adsorbed polymer as well as
their variation with changing force or strength of adhesion. All these
predictions appear in good agreement with our MC simulation results.

The polymer detachment transition under pulling is found to be of first order
whereby due to its dichotomic nature phase coexistence is impossible. This
absence of binodal states makes the polymer desorption under pulling a rather
unusual in comparison to conventional first-order phase transformation.

The critical line of desorption, while growing steadily when plotted in
dimensionless units of detachment force against surface potential, appears
``reentrant`` in absolute units of force against temperature. Thus, at very low
temperature the polymer is expected to be desorbed, with the growing $T$ it may
adsorb, and at even higher temperature - desorb again from the surface.

One finds that the crossover exponent, $\phi$, governing polymer
adsorption at criticality, whose exact value has been controversial for a
long time, depends essentially on interactions between different loops so that
$\phi$ may only vary within the limits $0.39 \le \phi \le 0.59$.

A point of more general importance for the statistical mechanics in general and
theory of phase transitions in particular is the issue of ensemble equivalence.
The latter implies an identity of the equation of state, regardless of which
statistical ensemble has been employed, whereas the fluctuations within the
different ensembles may be entirely different\cite{SKB}. For finite polymer
lengths, however, differences in the equation of state may also be visible. As
far as in practice one deals with finite polymer chains in laboratory
experiments, this difference is expected to be clearly manifested in cases of
practical concern.

{\em Acknowledgments}
We are indebted to A. Skvortsov, L. Klushin, J.-U. Sommer, and K. Binder for
useful discussions during the preparation of this work. A.~Milchev thanks the
Max-Planck Institute for Polymer Research in Mainz, Germany, for hospitality
during his visit in the institute. A.~Milchev and V.~Rostiashvili acknowledge
support from the Deutsche Forschungsgemeinschaft (DFG), grant No. SFB 625/B4.

\end{document}